\documentclass[showpacs,amsmath,amssymb,preprint]{revtex4}

\usepackage{graphicx,multirow}

\begin{document}
\title{An embedded-atom method model for liquid Co, Nb, Zr and supercooled binary alloys}
\author{Pascal Thibaudeau}
\affiliation{Commissariat \`a l'Energie Atomique\\ Le Ripault, BP 16, F-37260 Monts, France}
\author{Julian D. Gale}
\affiliation{Nanochemistry Research Institute\\ Department of Applied Chemistry, Curtin University of Technology, GPO Box U1987, Perth 6845, Western Australia}
\date{\today}

\begin{abstract}
The parameters of many-body potentials for Co, Nb and Zr metals, based on the embedded-atom method, have been systematically derived. The analytical potential scheme allows us to reproduce correctly the cohesive energies and structural properties of the pure metals and selected alloys making use of a small set of parameters. With a pair potential going smoothly to zero for a sufficient cutoff radius, radial partial and bond angular distribution functions for Co, Nb, Zr and alloys are computed using molecular dynamics simulations that ensure good quantitative agreement with the available experimental data up to the melting point. Atomic short range order is analysed in the light of consecutive Gaussian function decomposition and Honeycutt-Andersen indices.
\end{abstract}


\maketitle

%
\section{Introduction}
%
Today we are not surprised that a non-crystalline solid orders magnetically. It is known that, with few important exceptions, the amorphous and crystalline phases of the same material do not differ very much magnetically. In the last three decades, the discovery and large scale investigation of rapidly solidified alloys have made possible a new scenario in basic and applied magnetism. The unusual behaviour of the bulk metallic magnetic glasses (BMMGs) - formed by supercooling the liquid state of certain metallic magnetic alloys - usually occurs for systems containing atoms that exhibit a well-known sensitivity to the immediate neighbourhood \cite{Inoue:2003,Yavari:2007}. However, despite the large number of technological applications of BMMGs, the detailed origin of the links between structural and magnetic properties has yet to be established, but is strongly dependent on the atomic short-range order \cite{Miracle:2004}.

Nowadays, atomic-scale simulations of solids based on interatomic potentials are routinely performed to explore the short-range order of BMMGs \cite{Rozler:2000,Wolf:2004} as a precursor to the study of the magnetic ordering, which is evidently beyond such methods. However, the construction of realistic $n$-body potentials is mandatory to any simulations. When applied to metals and alloys, several major methods and extensions of construction of such potentials have been established from density functional theory, i.e., the embedded-atom method (EAM) \cite{Daw:1984} or tight-binding second moment approximation, i.e., the Finnis-Sinclair (FS) model \cite{Finnis:1984} and related models \cite{Hausleitner:1992,Philips:1994}. The two models have very similar computational requirements, and the names are often used interchangably, however there are some distinctions which come to the fore when considering multicomponent alloys \cite{Finnis:2003}. Because of the parameters involved in these model potentials, the EAM method was first used to study simple metals in their relevant crystalline structures \cite{Oh:1988,Johnson:1989}. Extensions to binary alloys were formulated \cite{Johnson:1989a} with special attention to hcp-fcc \cite{Cai:1996} and hcp-bcc \cite{Zhang:2005} systems. Because they are often based on spin-less approximations of electronic density, such effective model potentials traditionally neglect the magnetic ordering \cite{Finnis:1984,Finnis:2003,Ackland:1997,Mendelev:2003}. Notable exceptions have been published recently by Dudarev {\it et al.} \cite{Dudarev:2005} and Ackland {\it et al.} \cite{Ackland:2003,Ackland:2006}, but such methods are still in their infancy. Even in spin-less schemes, these effective potentials may be able to represent the compositional ordering which is, at least in the localised magnetism picture, a prerequisite for understanding the magnetic behaviour. Existing magnetic potentials concentrate on the one-site magnetism, for which the energetics reduces in form to a simple embedded-atom-type potential, explaining the success of standard EAM schemes on magnetic materials.

Supercooled Co$_{1-x}$Nb$_{x}$ and Co$_{1-x}$Zr$_{x}$ magnetic metallic glasses in the cobalt-rich region represent test cases for both hcp-bcc and hcp-hcp metal sub-systems. These compounds have been experimentally studied because they are strong ferromagnets, as revealed by a high value of the exchange constant \cite{Suran:1990}. Consequently, large short-range compositional inhomogeneities should induce significant variations of the long-range magnetic order. The present work considers both the construction of embedded-atom potentials for such materials, as well as the application to molecular dynamics (MD) simulations in order to assess one of the issues in the large field of these BMMGs, namely short-range order quasicrystallinity.
%
\section{Methodology for potentials}
%
In the FS method, the total internal energy $E$ of a $N$-atom system and the electron density, $\rho({\bf{R}}_i)$, for an atom located at ${\bf{R}}_i$ due to all other atoms are given as;
\begin{equation}
E=\frac{1}{2}\sum_{i,j,(i\neq j)}^N\phi(r_{ij})-\sum_{i=1}^NF(\rho({\bf{R}}_i)),\label{eam1}
\end{equation}
\begin{equation}
\rho({\bf{R}}_i)=\sum_{j\neq i}^Nf(r_{ij})\label{eam2},
\end{equation} 
where $f(r_{ij})$ is the electron density at atom $i$ due to atom $j$ as a function of the distance between them, $r_{ij}=\|{\bf{R}}_i-{\bf{R}}_j\|$ is the separation distance between atoms $i$ and $j$, $F(\rho({\bf{R}}_i))$ is the energy to embed atom $i$ in an electron density $\rho({\bf{R}}_i)$, and $\phi(r_{ij})$ is a two-body potential between atoms $i$ and $j$. As long as an angular independent formulation is considered, the electron density is a radial function only. For an alloy model, an embedding function, $F$, has to be specified for each atomic species supplemented by an atomic electron-density function, $f$, and a two-body potential, $\phi$, specified for each possible combination of atomic species. For uncompressed metals, Gupta \cite{Gupta:1981} and Tom\`anek {\it et al.} \cite{Tomanek:1985} have shown that the host electron density can be represented as an exponentially decreasing function of the distance to better account for atomic relaxation near impurities and surfaces. In this approximation, $f$ is given as;
\begin{equation}
f(r)=f_e\exp(-\chi(r/r_e-1)),\label{feam}
\end{equation}
where $f_e$ is a scaling factor determined by the cohesive energy, $E_c$, and the atomic volume, $r_e$ is the nearest-neighbour distance in the relevant pair of atoms and $\chi$ is an adjustable parameter. Analysing the interatomic interactions in effective-medium theory, Jacobsen {\it et al.} \cite{Jacobsen:1987} have shown that if an exponential form is chosen for the density function, then the interatomic potential, $\phi$, should also be an exponential function of the distance. In this study, the interatomic potentials of all the pairs considered are defined by a potential of the form;
\begin{equation}\label{phieam}
\phi(r)= \left\{
\begin{array}{ll}
A\exp(-r/r_0) & 0\le r\le r_1, \\
\displaystyle{\sum_{i=0}^5a_ir^{i}} & r_1\le r\le r_{m}, \\
0 & r_m\le r,
\end{array}
\right.
\end{equation}
where the interaction is designed so as to go smoothly to zero at the distance $r_{m}$ according to a polynomial spline function. The potentials are constructed subject to the constraints that the radial functions and their first and second derivatives must be continuous at the boundary points, and also that the function must have a stationary point at $r_{m}$. Once $A$, $r_0$, $r_1$ and $r_m$ are fixed, this procedure ensures that the coefficients, $\{a_i\}$, are uni\-quely determined by solving a simple $6\times 6$ linear system of equations. These coefficients are reported in Table \ref{cnzlib} for completeness. For all the pairs of atoms, $r_1=2.5${\AA} is kept fixed and corresponds to a typical radius where the stiff repulsive part of the potential ends in metals \cite{Dai:2006,Yifang:1996,Hausleitner:1990}. Because of the screening in metals, the stationary point is located at least between the second and third nearest neighbours for the lowest energy crystal phases as previously noted \cite{Bangwei:1993}. Since the embedding energy is assumed to be independent of the source of the electron density and the hopping integrals are a function only of a radial distance between atoms, the embedding functional $F$ is taken as 
\begin{equation}
F[\rho(r)]=\sqrt{\rho(r)}.
\end{equation}
This functional form gives a band energy proportional to the square root of the second moment of the electron density of states \cite{Gupta:1981}. However, moments of higher order cannot be expressed in such a simple analytical form and a more complex method must be applied \cite{Turchi:1985}. Johnson \cite{Johnson:1989a} has considered that since the electron density at any location is taken as a linear superposition of atomic electron densities, this function should be taken directly from monoatomic models with a relative scaling factor between elements for an alloy model. On the other hand, Finnis and Sinclair \cite{Finnis:1984} and Cleri and Rosato \cite{Cleri:1993} have considered mixed pair electron-density functions not necessarily connected to the atomic ones, removing the alloy scaling factors. 

For hcp-Co and hcp-Zr, the parameters for the atomic electron-density and the interatomic potential are fitted in order to reproduce the experimental cohesive energies, the unit cell parameters and the five independent elastic constants of these systems as given by Cleri and Rosato \cite{Cleri:1993}. Moreover, cell parameters and elastic constants of fcc-Co and bcc-Zr are also included during the fitting procedure as taken from references \cite{Yoo:1998,Modak:2006,Willaime:1991} and references therein. Generally, for a small cutoff distance the largest number of interacting neighbours per atom in the crystalline structure leads to the more stable phase. Ducastelle \cite{Ducastelle:1972} has shown that in the second-moment approximation, with interactions restricted to the nearest neighbours, the cohesive energy for the hcp and fcc phases is the same and the $c/a$ ratio is equal to the ideal value. Hence, it is necessary to go up to at least the fourth-moment approximation to discriminate between the fcc and hcp phases and to give a value of $c/a$ different from $2\sqrt{2/3}$. In our case, since the $c/a$ ratios are not taken to be the ideal one and because the potentials have a very short range, the cutoff distance of atomic electron density should be larger than that of the potential. So for the electron density, the cutoff distance is taken to be $4.87${\AA}, thus including up to seven shells of neighbours within hcp-Co, and three for both hcp-Zr and bcc-Nb, which are all the stable phases for each pure element. A check is also performed to ensure that at least three shells of neighbours have been included for the high pressure/high temperature structures since this is necessary to keep the relative energies of each phases in the correct order \cite{Willaime:1991}. Moreover, it has been observed that in incorporating the elastic constants of the bcc-Zr phase in the database, the fitting of both the cell parameters and cohesive energy of the hcp-Zr is rather poor with this cutoff radius, so only the hcp-Co, fcc-Co, hcp-Zr and bcc-Nb elastic constants are included during the fit. For Nb, the cohesive energy, the lattice parameter and the three independent elastic constants of the bcc-Nb phase are taken from reference \cite{Pasianot:1991} and the theoretical fcc-Nb cell parameter and cohesive energy are also included \cite{Haglund:1993}. No elastic constants of the fcc-Nb were found to incorporate into the training set of observables.

The selection of the functional form taken in Eq.(\ref{eam1})-(\ref{eam2}) is extended to AB alloys based on the second-moment form that has been applied to Zr \cite{Willaime:1989,Willaime:1991}. The embedding function, atomic electron-density function and two-body potential are assumed to be of the same form as in Eq.(\ref{feam})-(\ref{phieam}), with $\phi_{AB}$ and $\phi_{BA}$ assumed to be equal. The alloy potentials and atomic electron-density functions are determined independently of the monoatomic counterparts if sufficient data are available. However, it is known that for equilibrium immiscible systems it is a challenging task to fit cross potentials, since there is often insufficient experimental data related to the respective alloy compounds. In order to circumnavigate this problem, density functional calculations have been performed on selected intermetallic structures using the Quantum-ESPRESSO package \cite{espresso}. For these calculations, non-local ultrasoft pseudo-potentials are employed in combination with a plane-wave basis set. The generalized-gradient approximation, as parametrized by Perdew, Burke and Ernzerhof \cite{Perdew:1996}, is selected for the exchange and correlation term. For the Brillouin zone sampling, a 12$\times$12$\times$12 Monkhorst-Pack mesh is used for the $k$-point summation in the self-consistent calculations \cite{Monkhorst:1976} for all the primitive cells, which leads to converged structural parameters to within 0.1\% of the cell parameter. Thus the lattice constants and cohesive energies of several Co-(Zr,Nb) and Zr-Nb crystalline structures reported in Table ({\ref{binaries}}) are obtained and then included in the fitting procedure for the Co-(Zr,Nb) and Zr-Nb cross potentials and atomic electron-density functions. The parameters of the fitted terms are listed in Table \ref{cnzlib}.

\begin{table*}[ht]
\begin{center}
\begin{tabular}{ccccccc}\hline\hline
&Co-Co&Nb-Nb&Zr-Zr&Co-Nb&Co-Zr&Nb-Zr\\\hline
$f_e$ (eV$^{2}$)&1.6862&11.0160&3.4022&37.6462&18.2157&39.0130\\
$\chi$&3.4513&5.9621&3.6880&3.8908&2.8733&5.5304\\
$r_e $({\AA})&2.4968&2.8579&3.2133&2.5879&2.7985&3.0591\\
$A$ (eV)&162418.11&2758.20&27693.54&1353.63&13168.79&3092.13\\
$r_0 $({\AA})&0.1770&0.3495&0.2613&0.3211&0.2322&0.3754\\
$r_m $({\AA})&3.5269&3.6450&3.5798&4.3151&3.4794&3.4939\\
$a_0$ (eV)&193.2032&622.3713&1066.3553&107.1854&399.1098&1720.4867\\
$a_1$ (eV.{\AA}$^{-1}$)&-301.8495&-937.6188&-1616.1544&-143.1635&-619.6420&-2777.0048\\
$a_2$ (eV.{\AA}$^{-2}$)&188.4275&572.3560&984.8131&76.5117&386.7728&1810.9619\\
$a_3$ (eV.{\AA}$^{-3}$)&-58.7182&-176.1599&-301.0688&-20.4017&-121.1298&-593.4294\\
$a_4$ (eV.{\AA}$^{-4}$)&9.1312&27.2357&46.1097&2.7101&19.0083&97.3551\\
$a_5$ (eV.{\AA}$^{-5}$)&-0.5668&-1.6874&-2.8270&-0.1434&-1.1944&-6.3808\\
\hline\hline
\end{tabular}
\caption{Potential and atomic electron-density parameters for (Co,Nb,Zr) systems\label{cnzlib}.}
\end{center}
\end{table*}
The overall fitting procedure is performed in two separate steps. First, the densities and potentials are derived for the simple metals. Once the corresponding parameters are obtained, the fit is applied to selected binary metals for cross-densities and potentials without altering the terms for the simple metals. The same cutoff radius is kept constant during all the steps and the calculations are performed within the GULP computer code \cite{GaleRohl:2003}.

In Table \ref{simplemetal}, a list of some basic physical properties as computed by the present set of potentials and the corresponding experimental values are shown for Co, Nb and Zr.
\begin{table*}[ht]
\begin{center}
\begin{tabular}{clcccccc}\hline\hline
&&hcp-Co&fcc-Co&hcp-Zr&bcc-Zr&bcc-Nb&fcc-Nb\\\hline
\multirow{2}{*}{a({\AA})} 
&Fitted&2.5065&3.5414&3.1865&3.5358&3.3138&4.2198\\
&Experiment$^{(1)}$&2.507&3.544&3.2317&3.574&3.3&4.23\\
\multirow{2}{*}{c({\AA})} 
&Fitted&4.0606&-&5.2035&-&-&-\\
&Experiment$^{(1)}$&4.0689&-&5.1476&-&-&-\\
{$E_c$}&Fitted&-4.402&-4.360&-6.192&-5.970&-7.577&-7.383\\
(eV/atom)&Experiment$^{(1)}$&-4.386&-&-6.167&-6.13&-7.57&-7.39\\
\multirow{2}{*}{$C_{11}$ (GPa)} 
&Fitted&315&255&150&103&245&101\\
&Experiment$^{(2)}$&319&242&154&104&245&-\\
\multirow{2}{*}{$C_{12}$ (GPa)} 
&Fitted&155&159&80&89&132&122\\
&Experiment$^{(2)}$&166&160&67&93&132&-\\
\multirow{2}{*}{$C_{44}$ (GPa)} 
&Fitted&78&128&33&70&28&25\\
&Experiment$^{(2)}$&82&128&36&38&28&-\\
\multirow{2}{*}{$C_{13}$ (GPa)} 
&Fitted&111&159&52&89&132&122\\
&Experiment$^{(2)}$&102&160&65&93&132&-\\
\multirow{2}{*}{$C_{33}$ (GPa)} 
&Fitted&373&255&177&103&245&101\\
&Experiment$^{(2)}$&373&242&172&104&245&-\\
\multirow{2}{*}{$C_{66}$ (GPa)} 
&Fitted&80&128&35&70&28&25\\
&Experiment$^{(2)}$&77&128&44&38&28&-\\
\hline\hline
\end{tabular}
\caption{\label{simplemetal}Physical properties for Co, Zr and Nb simple metals as fitted with a cutoff radius of 4.87{\AA}. $^{(1)}$ Cohesive energies and lattice parameters are taken from Kittel \cite{Kittel:1966}, $^{(2)}$ elastic constants are taken from Simmons and Wang \cite{Simmons:1971}.}
\end{center}
\end{table*}
The fit correctly reproduces the structures and properties of hcp-Co, fcc-Co, hcp-Zr and bcc-Nb. The absolute average percentage difference between calculation and experiment is found to be 0.6\%, 0.7\% and 8\%, for the cell parameters, cohesive energies elastic constants, respectively. However, the elastic constants of fcc-Nb produced a negative Young's modulus, as expected, in this excited locally unstable structural phase \cite{Craievich:1997}. The properties of bcc-Zr are reproduced with a sufficient accuracy, except for the $C_{44}$ elastic constant where the largest percentage error of 45\% occurs. The potentials fitted by Willaime {\it et al.} \cite{Willaime:1991} also exhibit such a discrepancy though with a much larger error. This may be corrected by relaxing the constraint of the square-root form of the embedding functional and considering more neighbours \cite{Johnson:1989} or by including explicit angularly dependent terms in the potentials at a cost of additional parameters \cite{Baskes:1994}.

In Table \ref{binaries}, a list of some basic physical properties fitted from these potentials and the corresponding {\it ab initio} calculated values for selected binary alloys is shown.
\begin{table*}[ht]
\begin{center}
\begin{tabular}{lllcc}\hline\hline
&\multicolumn{2}{c}{a ({\AA})}&\multicolumn{2}{c}{$E_c$ (eV/atom)}\\
Structure&Fitted&{\it ab initio}&Fitted&{\it ab initio}\\\hline
CoZr (B2)&3.1612$^{(1)}$&3.1753&-6.9368&-6.9850\\
Co$_2$Zr (C15)&6.8944&6.9040&-6.8659&-6.7946\\
Co$_3$Zr (L1$_2$)&3.7058&3.7189&-6.3484&-6.5120\\
CoNb (B2)&2.9502&3.0523&-8.3599&-8.1750\\
Co$_2$Nb (C15)&6.6038&6.7357$^{(2)}$&-7.5813&-7.6012\\
Co$_3$Nb (L1$_2$)&3.5233&3.6289&-7.2423&-7.2425\\
ZrNb (B2)&3.4380&3.4380&-8.6865&-8.6200\\
Zr$_2$Nb (C15)&7.9961&8.0208&-8.2544&-7.7621\\
Nb$_2$Zr (C15)&7.8805&7.8565&-9.1236&-9.1725\\\hline\hline
\end{tabular}
\caption{\label{binaries}Lattice parameters and cohesive energies for selected binary metals in their corresponding symmetry as computed with a cutoff radius of 4.87{\AA}. $^{(1)}$ experimental value of 3.181{\AA} \cite{Buschow:1982a}, $^{(2)}$ experimental value of 6.774{\AA} \cite{Pargeter:1967}. The B2, C15 and L1$_2$ classification is relative to the strukturbericht structural types classification \cite{Smithell:1983}.}
\end{center}
\end{table*}
The B$2$, C$15$ and L$1_2$ denominations are relative to the strukturbericht structural types classification \cite{Smithell:1983}, such as B2 is the CsCl structure type, C15 is the MgCu$_2$ structure type and L1$_2$ is the Cu$_3$Au type. The fitting procedure appears to correctly reproduce the {\it ab initio} derived phase stability order and lattice parameters. However, the absolute cohesive energies are reproduced to a lesser extent. This may be also improved consistently by increasing the cutoff radius on Zr and Nb electron density terms, which includes more neighbours in the total energy sums of Eq.(\ref{eam1}). 
%
\section{Application to simple metals}
\begin{figure*}[p]
\begin{center}
\vspace{10mm}
\resizebox{0.9\columnwidth}{!}{\includegraphics[scale=1.0]{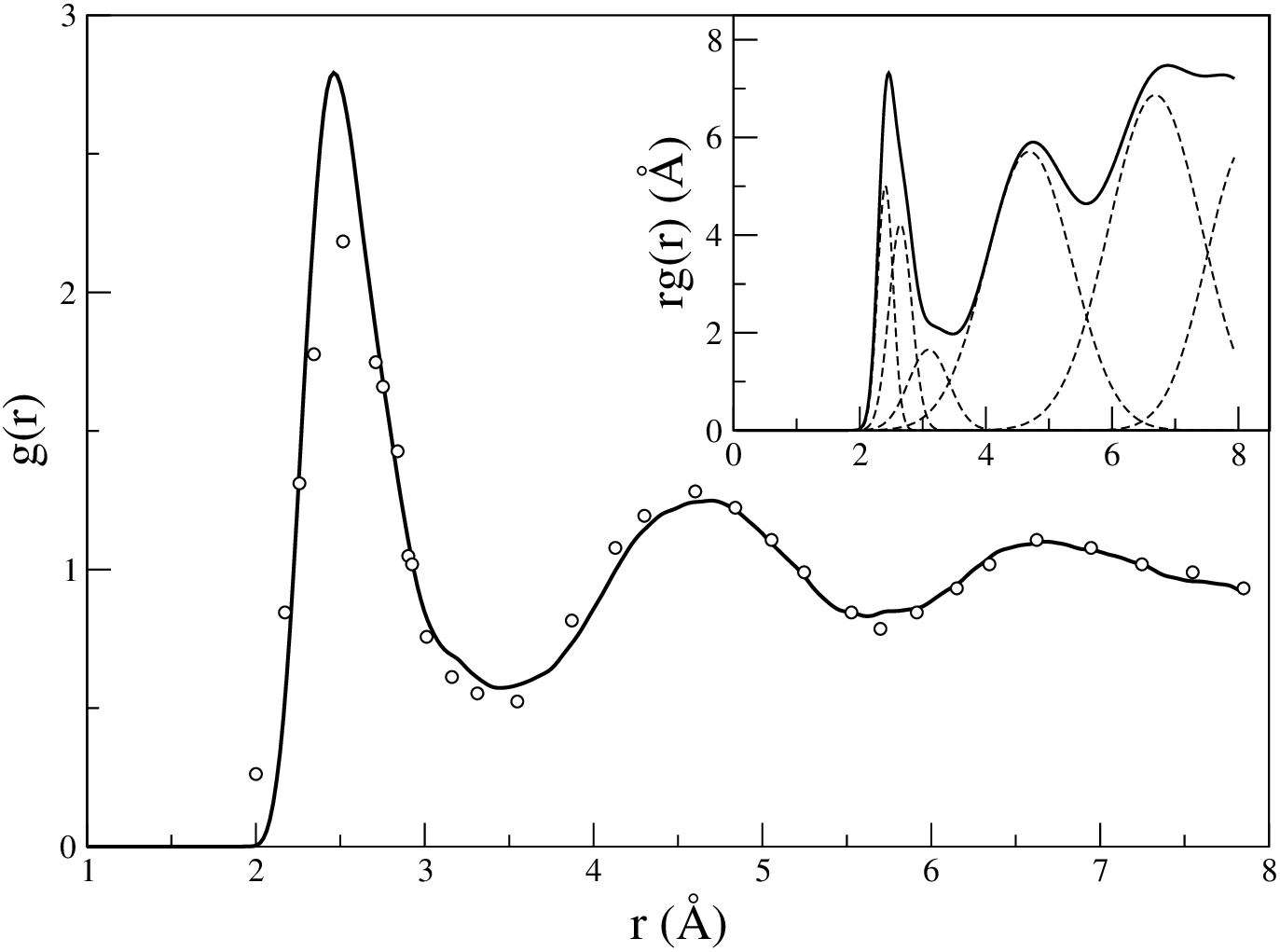}}
\caption{Simulated and experimental RPDF $g(r)$ of cobalt at a temperature of 1670 K, below the melting point. Experimental data from \cite{Holland-Moritz:2002} are shown by points. The insert shows the simulated $rg(r)$ at 1670 K and its analysis in Gaussian peaks (dashed curves). \label{coliq}}
\end{center}
\end{figure*}

%
The validity of this potential in describing the atomic interactions can be illustrated outside the original systems used for parametrisation by considering liquid cobalt. An MD simulation with 300 atoms, which allows an individual description up to the 13$^{th}$ nearest-neighbors in hcp-Co, and periodic boundary conditions was first performed within the isobaric, isothermal ensemble (NPT) at a temperature of 1670 K, which is slightly below the experimental melting point. The simulation was run for 200 ps with a time step of 0.1 fs to simulate the radial pair distribution function (RPDF). Once thermal equilibrium is reached, the RPDFs are sampled every 0.2 ps to produce an average. As these RPDFs are subject to statistical noise, a smoothing formula is used to replace each RPDF by a least-squares polynomial that fits a sub-range of several points. For all the calculated radial functions, a third-degree, five-point smoothing procedure is applied several times on the data until convergence \cite{Hildebrand:1965}. MD simulations are repeated on three different atomic configurations at the same temperature in order to sample more accurately the configurational space and an overall averaged RPDF is computed. This averaged RPDF compares very well to the experimental data \cite{Holland-Moritz:2002} as shown in Fig. \ref{coliq}.
The agreement is similar to that presented by Bhuiyan {\it et al.} \cite{Bhuiyan:1995} and more recently by Han and coworkers \cite{Han:2004} using much more elaborate EAM potentials, but in these works no quantitative analysis of the RPDFs was performed. However, the first peak of the RPDF $g(r)$ appears to be asymmetric and might be composed of more than one atomic shell. Following Kita {\it et al.} \cite{Kita:1994}, a decomposition of $rg(r)$ in consecutive Gaussian functions is applied. The insert in Fig. \ref{coliq} shows the decomposition of $rg(r)$ at 1670 K in six Gaussian functions where all 18 parameters are allowed to vary freely during the fit. The first coordination shell is defined by a cutoff distance $r_{c}$, which is taken to be the first minimum of $g(r)$. For this temperature, $r_{c}=3.46${\AA}. The first peak is composed of three Gaussian subpeaks located at $r_1=2.407${\AA}, $r_2=2.645${\AA} and $r_3=3.095${\AA}, respectively. The coordination number $N_{c}$ is calculated by integrating the Gaussian function according to $N_c=4\pi^{3/2}n_0A\sigma r_i$, where $n_0$ is the atomic density, $A$ is the amplitude of the Gaussian function, $\sigma$ is the square root of the variance, and $r_i$ is the maximum radius. For each subpeak, $N_c$ is equal to $3.46$, $4.71$ and $3.81$ with a sum of $11.98$. For a temperature of 1800K, which is slightly greater than the experimental melting point, the sum decreases to $11.57$. These coordination numbers are close to the experimental ones found in liquid Co \cite{Holland-Moritz:2002} ($12.5\pm 0.5$ at 1670 K and $12.1\pm 0.5$ at 1800 K) and consistent with those calculated for other metallic systems \cite{Kresse:1993}. Such a high value of the coordination number and the possibility of decomposing the first peak of $g(r)$ is an indication that the short-range order of the liquid Co is more complex than the one given by a simple icosahedral ordering as suggested by Holland {\it et al.} \cite{Holland-Moritz:2002}. Moreover, performing MD simulations for these two temperatures allows us to predict a variation of the density with the temperature of $d\rho/dT=-9.68$ $10^{-4}$gcm$^{-3}$K$^{-1}$ in good agreement with the experimentally reported value of $d\rho/dT=-9.88$ $10^{-4}$gcm$^{-3}$K$^{-1}$ value \cite{Smithell:1983}. 

\begin{figure*}[p]
\begin{center}
\vspace{10mm}
\resizebox{0.9\columnwidth}{!}{\includegraphics[scale=1.0]{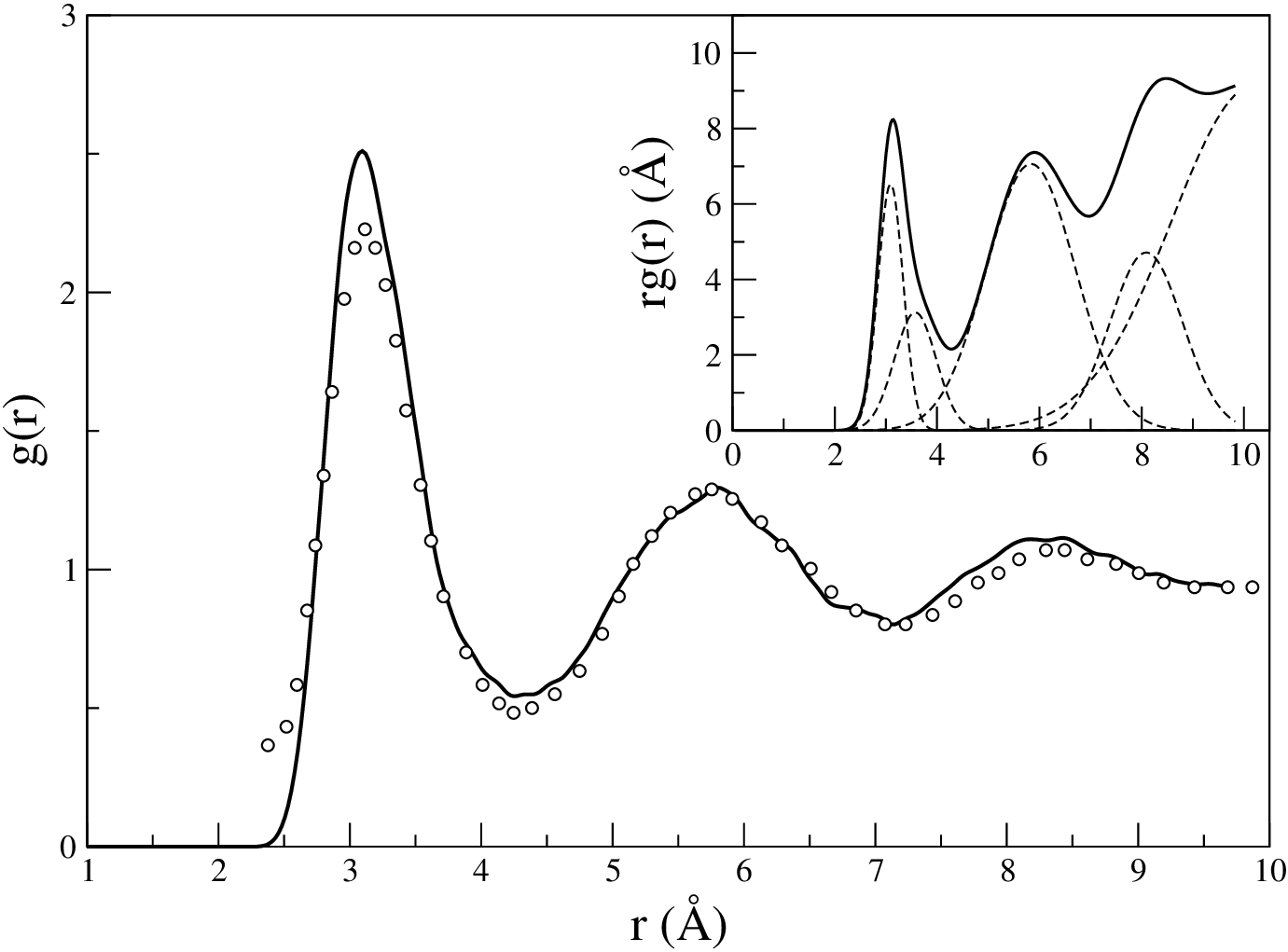}}
\caption{Simulated and experimental RPDF $g(r)$ of liquid zirconium at a temperature of 2290 K. Experimental data from \cite{Schenk:2002} are shown by points. The insert shows the simulated $rg(r)$ at 2290 K and its analysis in Gaussian subpeaks (dashed curves). \label{zrliq}}
\end{center}
\end{figure*}

MD simulations were repeated under the same conditions for pure Zr to a higher temperature of 2290 K, above the experimental melting point. The experimental RPDF is compared against the simulation of this liquid state, shown in Fig. \ref{zrliq}. For this metal at that temperature, the first peak is composed of two Gaussian functions located at $r_1=3.091${\AA} and $r_2=3.574${\AA}, respectively. The ratio $r_2/r_1=1.156$ is close to that of the two first nearest-neighbour distances for a bcc lattice $2/\sqrt{3}=1.1547$. The coordination number $N_{c}$ for each subpeak is $6.10$ and $5.59$ with a sum of $11.69$. This value is close to the experimental one of $11.9\pm 0.5$ found in liquid Zr \cite{Schenk:2002}. Even if the ratio of the radii tends to favour a bcc lattice, the corresponding coordination is very different. 

\begin{figure*}[p]
\begin{center}
\vspace{10mm}
\resizebox{0.9\columnwidth}{!}{\includegraphics[scale=1.0]{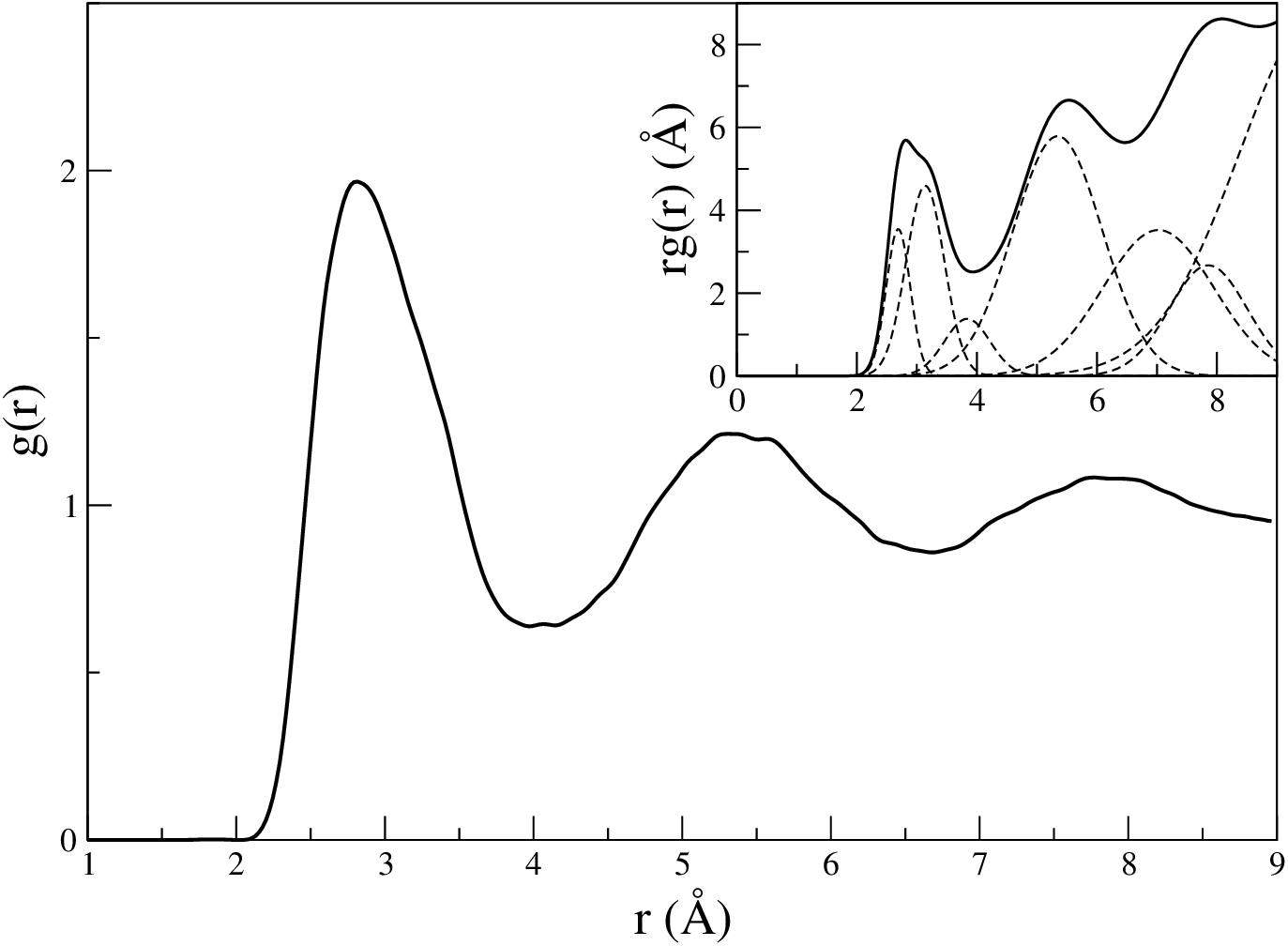}}
\caption{Simulated RPDF $g(r)$ of liquid niobium at a temperature of 2750 K. The insert shows the simulated $rg(r)$ and its analysis in seven Gaussian subpeaks (dashed curves). \label{nbliq}}
\end{center}
\end{figure*}

MD simulations are repeated under the same conditions for pure Nb to a higher temperature of 2750 K, which is the experimental melting point. The simulated RPDF of this liquid state is shown in Fig. \ref{nbliq}. For this metal, the first peak of $rg(r)$ is strongly asymmetric and is composed of three Gaussian functions up to $r_{c}=4.0${\AA}, located at $r_1=2.691${\AA}, $r_2=3.141${\AA} and $r_3=3.840${\AA}. The ratio $r_2/r_1=1.167$ is greater than that of the two first nearest-neighbour distances for a bcc lattice. The coordination number $N_{c}$ for each subpeak is $2.95$, $7.06$ and $3.10$ with a sum of $13.11$. To our knowledge, no experimental data on the radial distribution function of pure Nb is available in order to compare with.

\begin{figure*}[p]
\begin{center}
\vspace{10mm}
\resizebox{0.9\columnwidth}{!}{\includegraphics[scale=1.0]{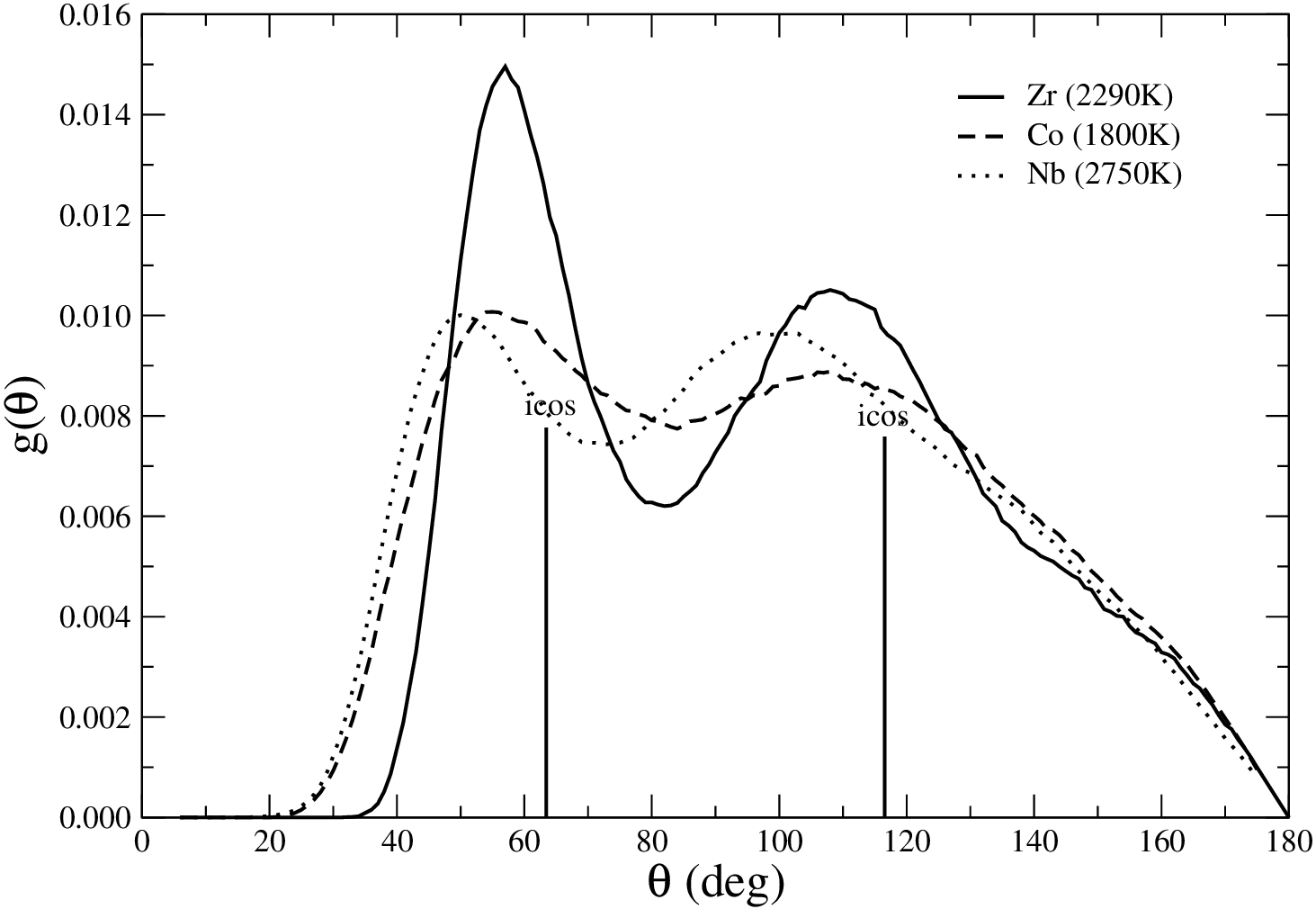}}
\caption{Bond angle distribution at $T=2290$ K for liquid Zr (solid line), at $T=1800$ K  for liquid Co (dashed line) and at $T=2750$ K for liquid Nb (dotted line). The peaks in the bond angle distribution for perfect icosahedral order are indicated by the vertical lines. \label{bondangle}}
\end{center}
\end{figure*}

The short-range order can also be examined by calculating the bond angle distribution functions $g(\theta)$ that represent the angle between the bonds connecting a central atom to two neighbouring atoms, as illustrated in Fig. \ref{bondangle} for liquid cobalt, zirconium and niobium. The angle is calculated for pairs of interatomic distances given by a cutoff corresponding to the first minimum of the RPDF (i.e. 3.5{\AA} for Co, 4.3{\AA} for Zr and 4.1{\AA} for Nb). In the case of Zr, the calculated distribution exhibits a prominent peak near $\theta=57^{\circ}$ (close to an equilateral triangle), a broader maximum near $\theta=109^{\circ}$ and a rather flat maximum near $\theta=150^{\circ}$. In the case of Co, the first peak is broader and close to $55^{\circ}$ whereas the second peak enlarges but remains at $109^{\circ}$. For Nb, the situation is different with two broader peaks near $50^{\circ}$ and $99^{\circ}$. Viewing the structure in terms of dominant clusters, the bond angle with the highest density at the nearest neighbour distance are for a regular icosahedron $63.4^{\circ}$ and $116.4^{\circ}$, while for fcc the prominent angles are $60^{\circ}$, $90^{\circ}$ and $120^{\circ}$. For hcp, angles of $109.471^{\circ}$ and $146.443^{\circ}$ are added when the ratio $c/a=2\sqrt{2/3}$ but they are less frequent. For a bcc lattice, the prominent angles are $70.53^{\circ}$ and $109.471^{\circ}$. In the Zr structure, the first peak tends to favour the fcc and hcp structures while the second peak tends to favour the hcp and bcc. This means that the dominant structure should be hcp. However, the angles of $90^{\circ}$ and $120^{\circ}$ are not so strong whereas some defective icosahedron angles should be there too. This suggests a predominantly distorted icosahedral character. For the same reasons, the case of Co also favours defective icosahedron as well. For Nb as the first peak is much more located near $50^{\circ}$, this suggests much more intriguing short-range structures with less neighbours.

To assess more quantitatively the local structures in amorphous alloys, Honeycutt and Andersen (HA) analysis has been proven to successfully differentiate face-centered cubic, hexagonal close packed, icosahedron and binary bcc structures \cite{Honeycutt:1987,Jonsson:1988}. To perform such analysis, a set of four indices is constructed for each pair: (i) the first index denotes to what peak of the RPDF, $g(r)$, the pair under consideration belongs; (ii) the second index represents the number of near neighbours shared by this pair; (iii) the third index counts the number of nearest-neigbour bonds among the shared neighbours; (iv) and a fourth index is used to differentiate configurations with the same three indices, but with a different topology. For instance, fcc crystals are fully described by four pairs, such as 1421, 2101, 2211 and 2441, whereas hcp crystals also contains the 1422 and 2331 pairs in addition. Moreover, the 1441, 1661, 2101, 2211, and 2441 are the only pairs in a perfect bcc crystal. Icosahedral order is described by Mackay icosahedra, composed of twinned, distorted fcc tetrahedra with an index of 1551, whereas the 1541 and 1431 indices are more characteristic of a distorted icosahedral local order. Up to the distance cutoff corresponding to first minimum of each $g(r)$, Table \ref{HA_simple_liq} reports the HA analysis in the liquid state of each simple elements.
\begin{table}[ht]
\begin{center}
\begin{tabular}{lccc}\hline\hline
index&Co&Zr&Nb\\
\hline
1311&0.07$\pm$0.01&0.06$\pm$0.01&0.03$\pm$0.01\\
1321&0.07$\pm$0.01&0.06$\pm$0.01&0.05$\pm$0.01\\
1421&0.04$\pm$0.01&0.03$\pm$0.01&0.02$\pm$0.01\\
1422&0.07$\pm$0.01&0.06$\pm$0.01&0.04$\pm$0.01\\
1431&0.19$\pm$0.01&0.17$\pm$0.01&0.13$\pm$0.01\\
1441&0.04$\pm$0.01&0.05$\pm$0.01&0.07$\pm$0.01\\
1541&0.15$\pm$0.01&0.15$\pm$0.01&0.11$\pm$0.01\\
1551&0.12$\pm$0.01&0.14$\pm$0.01&0.14$\pm$0.01\\
1661&0.05$\pm$0.01&0.06$\pm$0.01&0.07$\pm$0.01\\
\hline
2101&1.58$\pm$0.02&1.55$\pm$0.01&1.52$\pm$0.02\\
2211&0.95$\pm$0.02&0.94$\pm$0.02&0.96$\pm$0.02\\
2321&0.23$\pm$0.01&0.22$\pm$0.01&0.21$\pm$0.01\\
2331&0.50$\pm$0.02&0.53$\pm$0.02&0.58$\pm$0.02\\
2441&0.08$\pm$0.01&0.08$\pm$0.01&0.07$\pm$0.01\\
\hline\hline
\end{tabular}
\caption{\label{HA_simple_liq}Honeycutt and Andersen analysis of the simulations in the liquid state for Co (T=1670K), Zr (T=2290K) and Nb (T=2750K).}
\end{center}
\end{table}
The microscopic analysis emerging from the data of Table \ref{HA_simple_liq} indicates that the short-range order of the liquid state is dominated by distorted icosaheral and icosahedral structures since the 1541, 1431 and 1551 indices respectively are large as anticipated. The high value of the 2331 pairs is also an indication of the icosahedral order. The small distortion from perfect icosahedral order observed in the angular distributions of these liquids suggests that the local icosahedral order should dominate. However, small distortion form a perfect tetrahedron does not form different HA indices from those for a perfect icosahedron. Our HA analysis shows that the icosahedral distortion is larger than reported by the bond angle distribution curves. This result is in agreement with the experimental investigation on liquid Ti, Zr and Ni conducted by Kim and Kelton \cite{Kim:2007}. Using first-principles molecular dynamics simulations, Jakse and Pasturel \cite{Jakse:2003} have concluded there exists competition between a polyhedral and bcc-type short-range order in liquid and supercooled Zr, where\-as Kim and Kelton \cite{Kim:2007} have reported no regular dominant cluster type that can describe the experimental liquid structure of transition metals, including Zr. This is supported by the values of the HA indices reported in Table \ref{HA_simple_liq} which are not very different from each element in the liquid state. The abundance of the 1661 pairs indicates that bcc order is very low, but slightly increases when going from Co to Nb. Interestingly, the lowest energy geometrical structures of magnetic cobalt clusters mainly follow an icosahedral growth pattern with some cubic-type structures at some particular sizes \cite{Rodriguez-Lopez:2004}.

For liquid Zr, these low values have been reported both experimentally \cite{Kim:2007} and using first-principles molecular dynamics simulations \cite{Jakse:2003}. Furthermore for liquid Zr, Jakse and Pasturel have performed HA analysis on the inherent structures and found an abundance of the 1551 pairs in the liquid state that is twice the value found here. Such structures indicate the presence of perfect icosahedra as a local minima of the potential energy surface. For instance on liquid Nb, the 1551 index goes to 0.30$\pm$0.01 on inherent structure.
%
\section{Application to binary alloys}
\begin{figure*}[p]
\begin{center}
\vspace{10mm}
\resizebox{0.9\columnwidth}{!}{\includegraphics[scale=1.0]{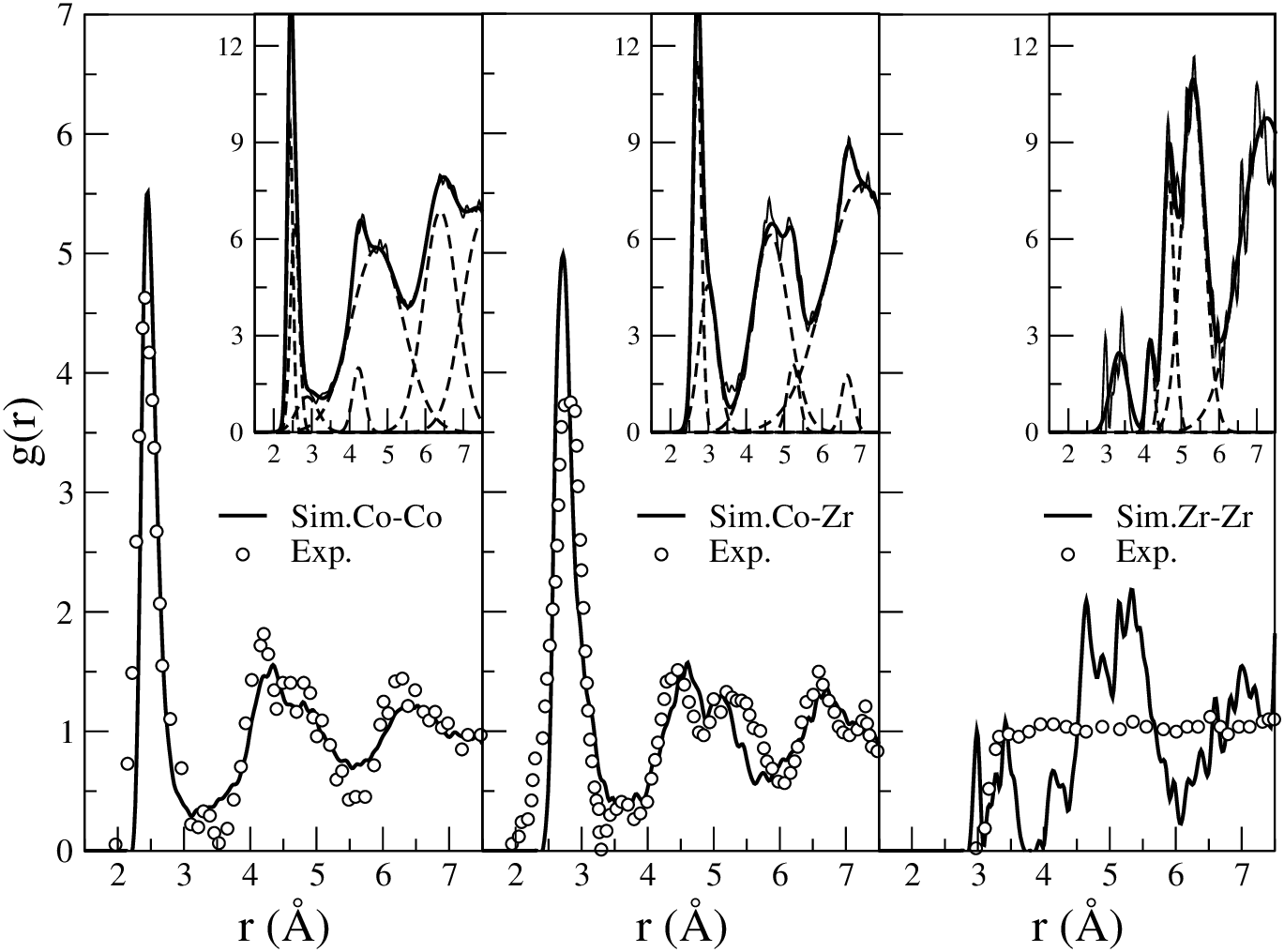}}
\caption{Radial partial distribution functions of Co$_{0.9}$Zr$_{0.1}$ at 300 K compared with experiment (open circles) \cite{Babanov:1998}. For each atomic pair, the insert shows the simulated $rg(r)$ at 300 K and its analysis in Gaussian peaks (dashed curves).\label{gijcozr9010}}
\end{center}
\end{figure*}

%
To validate the quality and transferability of our potentials, lattice constants and atomic internal positions of several crystal structures not entering in the fit have been calculated minimising the free energy at T=300K, and compared with the experimental values \cite{Shurin:1965,Kuzma:1966}. The results are summarised in Table \ref{verif} for the varying degrees of freedom according to the corresponding space group and the agreement is generally good with an average absolute error of 1.76\%. 
\begin{table*}[ht]
\begin{center}
\begin{tabular}{llllll}\hline\hline
&\multicolumn{2}{c}{Co$_7$Nb$_6$}&\multicolumn{2}{c}{Co$_{23}$Zr$_6$}\\\hline
&exp.\cite{Shurin:1965}&EAM(300K)&&exp.\cite{Kuzma:1966}&EAM(300K)\\
a({\AA})&5.01&5.019(+0.18\%)&a({\AA})&11.516&11.484(-0.28\%)\\
c({\AA})&26.5&24.66(-6.95\%)&x$_{Co_3}$&0.378&0.3791(+0.30\%)\\
x$_{Co_1}$&0.5&0.5005(+0.10\%)&x$_{Co_4}$&0.178&0.1732(-2.71\%)\\
y$_{Co_1}$&0.5&0.4995(-0.10\%)&x$_{Zr_1}$&0.208&0.2075(-0.26\%)\\
z$_{Co_1}$&0.59&0.5801(-1.67\%)&&&\\
z$_{Nb_1}$&0.167&0.1664(-0.34\%)&&&\\
z$_{Nb_2}$&0.346&0.3211(-7.19\%)&&&\\
z$_{Nb_3}$&0.448&0.4359(-2.71\%)&&&\\
\hline\hline
\end{tabular}
\caption{\label{verif}Lattice constants and atomic internal positions of Co$_7$Nb$_6$ and Co$_{23}$Zr$_{6}$ calculated with our potentials at 300K and compared to the experimental values \cite{Shurin:1965,Kuzma:1966}.}
\end{center}
\end{table*}
To complete the validation, the elastic constants of CoNb and Co$_3$Nb are calculated {\it ab initio} applying finite differences to the stress tensor and compared with those calculated analytically by our potentials. The results are shown in Table \ref{elastic} and the agreement is satisfying. 
\begin{table}[ht]
\begin{center}
\begin{tabular}{lcccc}\hline\hline
&\multicolumn{2}{c}{CoNb ($D_2$)}&\multicolumn{2}{c}{Co$_3$Nb ($L_{12}$)}\\\hline
&{\it ab initio}&EAM&{\it ab initio}&EAM\\
C$_{11}$ (GPa)&251&242&368&357\\
C$_{12}$ (GPa)&173&143&164&194\\
C$_{44}$ (GPa)&60&71&160&131\\\hline\hline
\end{tabular}
\caption{\label{elastic}{\it ab-initio} elastic constants of selected binaries as compared to the present set of potentials.}
\end{center}
\end{table}
As for liquids, MD simulations are performed on Co$_{0.9}$Zr$_{0.1}$ for a 300 atom system with periodic boundary conditions. First, the atoms are placed randomly into the simulation cell using a hard sphere criterion based on their atomic radii and the cell volume is then adjusted according to the phenomenological Miedema theory \cite{Miedema:1980}, which gives good estimates for the experimental volume of glasses of these alloys. Then at constant pressure and a temperature of 1800 K (higher than the liquidus phase boundary for this alloy composition), MD has been run for 200 ps to simulate the liquid phase. The sample is then quenched at a rate of $7.5\times 10^{14}$Ks$^{-1}$ and maintained at 300 K for at least a further 200 ps. In Fig. \ref{gijcozr9010}, the simulated RPDFs are computed and compared against experiment \cite{Babanov:1998}.
The simulated partial distribution functions for cobalt compare well to experiment including the position of the first and second peaks. However, the experimental Zr-Zr RPDF does not exhibit a structural trend, whereas our simulation does. R{\"o}{\ss}ler and Teichler have reported similar results in their study of atomic mobilities and structural properties of supercooled amorphous Co$_{1-x}$Zr$_x$ using very different interatomic potentials \cite{Rozler:2000}. This suggests a possible lack of resolution both in X-ray diffraction and EXAFS because of a low contribution in the spectra of these minority atoms, as previously anticipated \cite{Rozler:2000}. However, our potentials seem to reproduced correctly the position of the first peak for Zr-Zr and the double peak character of the second peak in the Co-Co and Co-Zr distributions in comparison with ref.\cite{Rozler:2000}. The maximum of the Co-Co peak is simulated to occur at 2.44{\AA}, compared with an experimental result of 2.42{\AA} \cite{Babanov:1998}. Using the Gaussian subpeak analysis, this first peak is found to be composed of 3 shells at $2.43${\AA}, $2.55${\AA} and $2.88${\AA} with coordination numbers of $4.42$, $4.83$ and $2.30$. The total coordination number up to $r_{c}=3.12${\AA} is equal to $11.55$ in comparison with $10.90$ found experimentally. In our simulations a narrower scattering of the positions of the first 3 subpeaks is observed, in contrast to the liquid state, which is anticipated during the cooling. In the closest crystalline form, Co$_{23}$Zr$_{6}$, the highest coordination number is obtained for a pair of cobalt atoms located at $2.4334${\AA} in a cubic cluster and other local structures with $4$ and $3$ neighbours are also found at 2.36{\AA}, 2.51{\AA}, 2.51{\AA} and 2.81{\AA} \cite{Kuzma:1966}. This suggests tetrahedral clustering below $r_c$ or defective icosahedra up to $r_c$ in this supercooled alloy. On the other hand, the maximum of the first peak in Co-Zr is simulated to be at 2.79{\AA}, in excellent agreement with the reported experimental distance of 2.79{\AA} \cite{Babanov:1998}. The analysis through Gaussian functions reveals 2 subpeaks located at 2.71{\AA} and 3.00{\AA} with a coordination number of 8.68 and 8.40, respectively. This also suggests a clustering of the bcc-type at very short range.

\begin{figure*}[p]
\begin{center}
\vspace{10mm}
\resizebox{0.9\columnwidth}{!}{\includegraphics[scale=1.0]{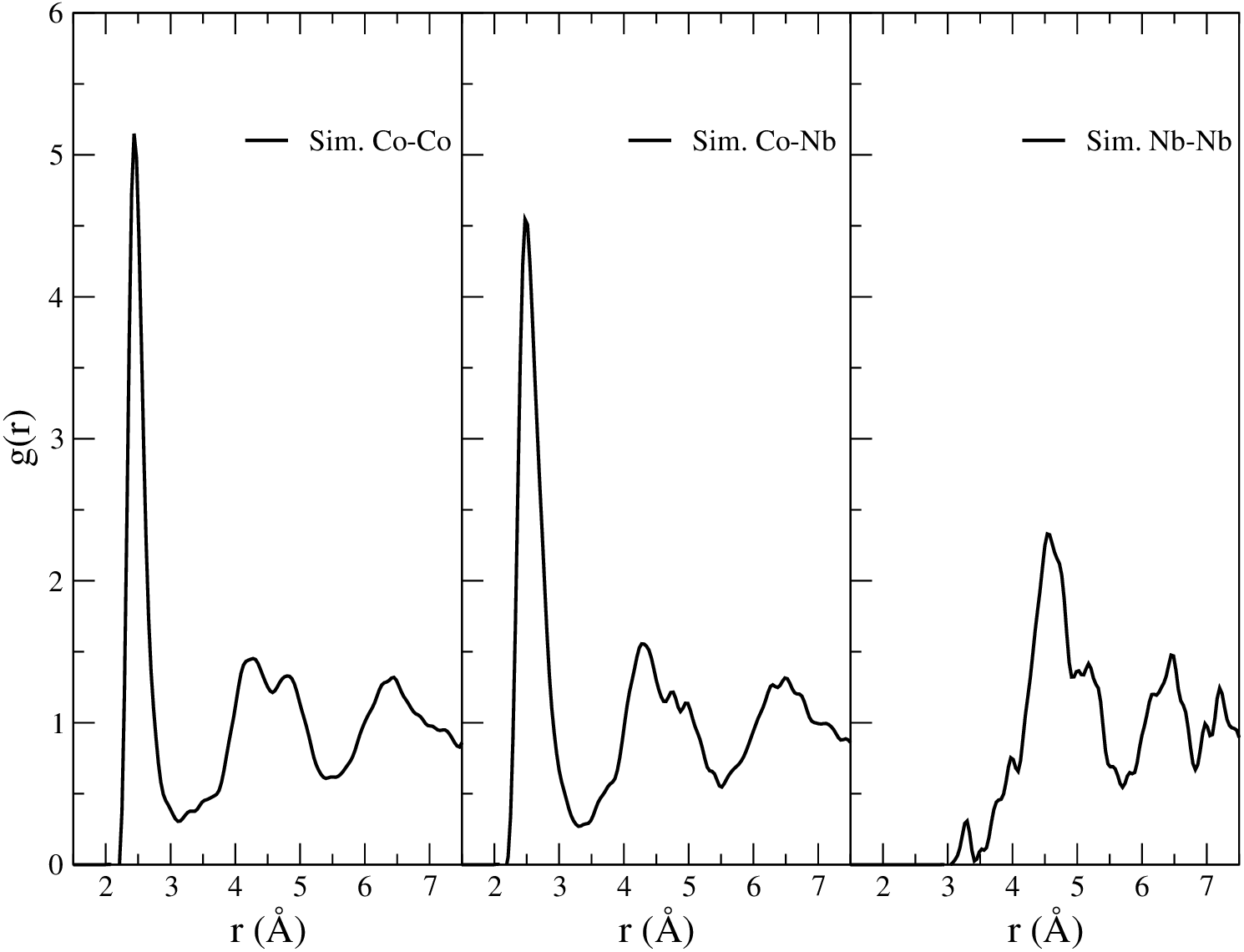}}
\caption{Simulated radial partial distribution functions of Co$_{0.9}$Nb$_{0.1}$ quenched at 300K.\label{rdfconb}}
\end{center}
\end{figure*}

\begin{figure*}[p]
\begin{center}
\vspace{10mm}
\resizebox{0.9\columnwidth}{!}{\includegraphics[scale=1.0]{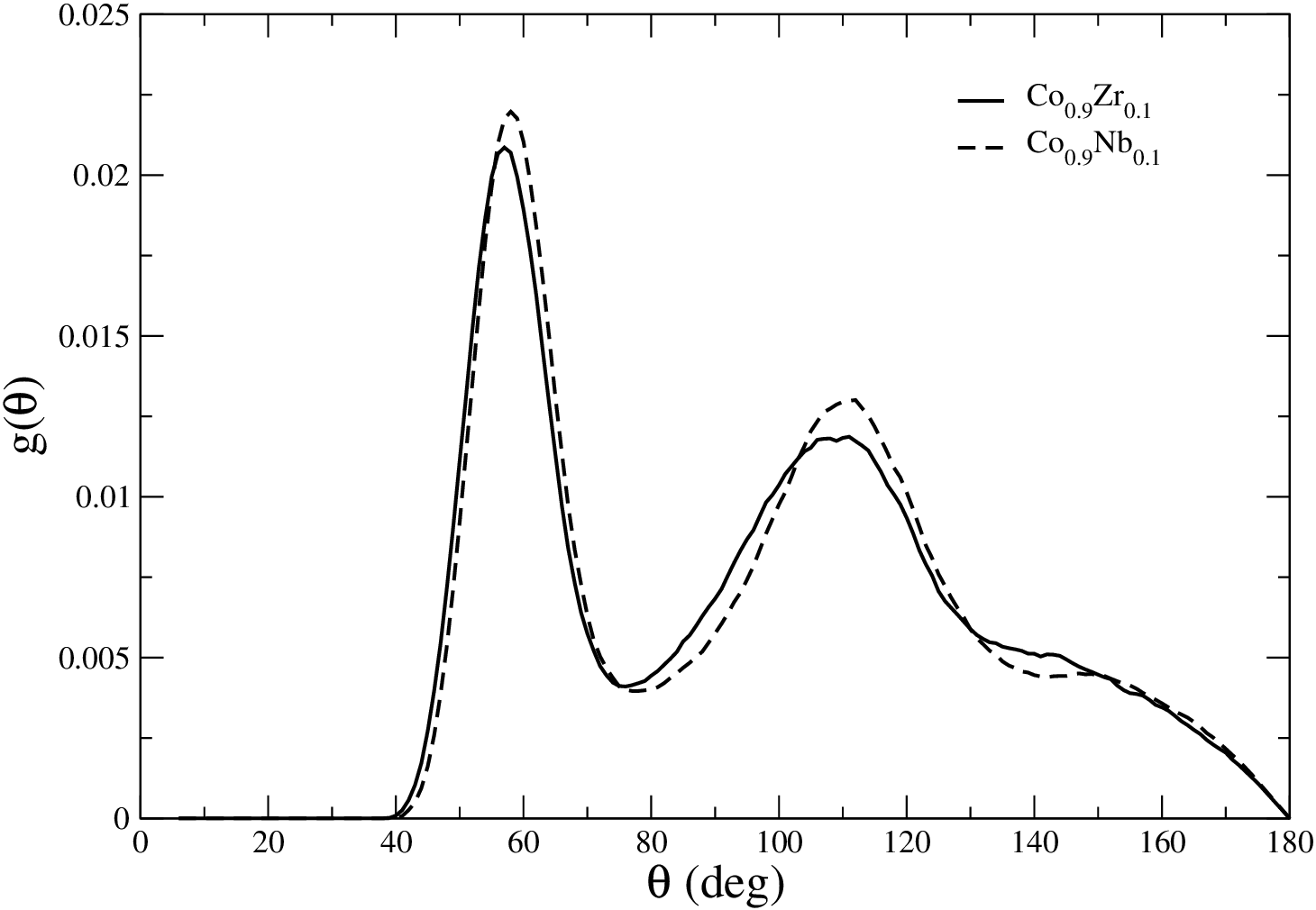}}
\caption{Bond angular distribution functions of Co$_{0.9}$Zr$_{0.1}$ and Co$_{0.9}$Nb$_{0.1}$ quenched at 300K.\label{theta_alloys}}
\end{center}
\end{figure*}

MD simulations have also been performed for Co$_{0.9}$Nb$_{0.1}$ under the same conditions. The reported RPDFs are shown in Fig. \ref{rdfconb}. In Co$_{0.9}$Nb$_{0.1}$ and Co$_{0.9}$Zr$_{0.1}$, the Co-Co first peak distance is calculated to be at 2.437{\AA} and is not affected by the non-magnetic added atoms at such a low concentration. The situation changes for the second and third peaks with a much more distinct third shell in Co$_{0.9}$Nb$_{0.1}$ than in Co$_{0.9}$Zr$_{0.1}$. Interestingly, the Co-Nb (resp.Co-Zr) first neighbour equilibrium distance is 2.49{\AA} (resp.2.79{\AA}). This is lower than the simple prediction related to their corresponding atomic radii (2.70{\AA} (resp.2.85{\AA})) \cite{Vainshtein:2000}. However, Jamet {\it et al.} \cite{Jamet:2000} have reported a Co-Nb distance of 2.58{\AA} in studying cobalt nanoparticles embedded in a niobium matrix.    
The simulated correlation in the minority pairs is structured in the Co-Nb system, as for Co-Zr. However, in both cases the first shell of neighbours seems to be depleted of their atoms to fill the second or third shells. It is doubtful that these structures are an artefact of the low number of atoms considered in our simulations because R{\"o}{\ss}ler and Teichler have simulated systems more than twice the size of ours and found the same behaviour. Up to $r_c=3.1${\AA}, the bond angular distributions of these two systems are calculated and shown in Fig. \ref{theta_alloys}. These distributions exhibit well-structured peaks suggesting more crystalline environments including a $150^{\circ}$ distinct angle in the Co$_{0.9}$Nb$_{0.1}$. As the concentration of minority atoms is low, these distributions are dominated by hcp-like Co clusters upon cooling. This trend is more pronounced for added Nb atoms than Zr atoms probably because of their corresponding atomic radii.  
To assess such hypothesis, the HA indices are calculated and reported in Table \ref{HA_cozrnb}. 
\begin{table}[ht]
\begin{center}
\begin{tabular}{lcc}\hline\hline
index&Co$_{0.9}$Zr$_{0.1}$&Co$_{0.9}$Nb$_{0.1}$\\
\hline
1311&0.02&0.06\\
1321&0.03&0.03\\
1421&0.02&0.06\\
1422&0.04&0.11\\
1431&0.14&0.21\\
1441&0.05&0.01\\
1541&0.15&0.17\\
1551&0.24&0.16\\
1661&0.08&0.03\\
\hline
2101&1.49&1.57\\
2211&0.81&0.87\\
2321&0.20&0.20\\
2331&0.70&0.62\\
2441&0.10&0.13\\
\hline\hline
\end{tabular}
\caption{\label{HA_cozrnb}Honeycutt and Andersen analysis of the simulations for supercooled Co$_{0.9}$Zr$_{0.1}$ and Co$_{0.9}$Nb$_{0.1}$ at 300K. The absolute error bars of the abundances are 0.01.}
\end{center}
\end{table}
In Co$_{0.9}$Zr$_{0.1}$ the dominant pairs are 1551 and 1541 exhibiting more icosahedral than distorted icosahedral order. This tendency is inverted in Co$_{0.9}$Nb$_{0.1}$ system with a larger amount of 1431 pairs favouring distorted icosahedral order. This is explained by a smaller difference in atomic radii between Co and Nb than Co and Zr because an atomic size difference of approximately 10\% can relieve spatial frustration and stabilise the icosahedral structure \cite{Nelson:1989}. In Co$_{0.9}$Nb$_{0.1}$ the 1422 pairs are abundant indicating an hcp order which is less present in Co$_{0.9}$Zr$_{0.1}$ system. Even if the value is rather low, the 1661 pairs are of importance indicating a slight bcc order, as anticipated. It is observed that while useful for gaining understanding of the evolution of the dominant short-range order, the single cluster model cannot capture the richness of the supercooled binary alloys structures.
%
\section{Conclusion}
%
A fitting procedure has been performed to consistently derive a self-consistent set of many-body parameters for Co, Zr and Nb simple metals and selected alloys, including validation against first principles results where there are gaps in the experimental data. Combined with MD simulations, these parameters allow us to calculate RPDFs and bond angular distributions in the liquid phase for Co, Zr and Nb. Applied to supercooled binary alloys, clear short-range order is shown in agreement with available experiments for the majority pairs. The situation is different for the minority pairs within the Co-rich region. In this region, both simulated Zr-Zr and Nb-Nb RPDFs are correlated and exhibit transfers of atoms from the first shells of neighbours to the second and third. The Honeycutt-Andersen analysis exhibits mainly both distorted and pure icosahedral orders of various degrees in competition with other crystalline orders in the liquid phases and supercooled alloys.
%
%

JDG would like to thank the Government of Western Australia for a Premier's Research Fellowship.
%
%
\bibliographystyle{unsrt}


\end{document}